\documentclass[showpacs, secnumarabic, amssymb, nobibnotes, aip, apl, reprint]{revtex4-1}
\usepackage{graphicx}%
\usepackage{dcolumn}%
\usepackage{bm}%
\usepackage{lineno}

\begin{document}

\title{Magnetically active vacancy related defects in irradiated GaN layers}

\author{L.~Kilanski}
 \altaffiliation[Also at ]{Institute of Physics, Polish Academy of Sciences, Warsaw, Poland} 
 \email{kilan@ifpan.edu.pl}
\author{F.~Tuomisto}
\affiliation{Department of Applied Physics, Aalto University, P.O. Box 11100, FI-00076 Aalto Espoo, Finland}

\author{R.~Szymczak}
\affiliation{Institute of Physics, Polish Academy of Sciences, Al. Lotnikow 32/46, 02-668 Warsaw, Poland}

\author{R.~Kruszka}
\affiliation{Institute of Electron Technology, Al. Lotnikow 46, 02-668 Warsaw, Poland}

\date{\today}

\begin{abstract}

We present the studies of magnetic properties of 2 MeV $^{4}$He$^{+}$-irradiated GaN grown by metal-organic chemical-vapor deposition. Particle irradiation allowed controllable introduction of Ga-vacancy in the samples. The magnetic moments with concentrations changing between 4.3$...$8.3$\times$10$^{17}$ cm$^{-3}$ showing superparamagnetic blocking at room temperature are observed. The appearance of clear hysteresis curve at $T$$\,$$=$$\,$5$\;$K with coercive field of about $H_{C}$$\,$$\approx$$\,$270$\;$Oe suggests that the formation of more complex Ga vacancy related defects is promoted with increasing Ga vacancy content. The small concentration of the observed magnetically-active defects with respect to the total Ga- vacancy concentration suggests that the presence of the oxygen/hydrogen-related vacancy complexes is the source of the observed magnetic moments.

\end{abstract}





\maketitle

\linenumbers

In spite of intense research activities for more than 40 years, the challenge of achieving room temperature ferromagnetism in semimagnetic semiconductors is till present.\cite{MacDonald2007a, Awschalom2007a} Usually, semimagnetic semiconductors (SMCS's) are designed on a basis of III-V or II-VI semiconducting matrix with dilution of a few percent of transition metal or rare earth ions possessing unfilled 3d or 5f shells.\cite{Furdyna1998a} In addition to many reports of high temperature magnetic ordering in inhomogeneous SMCS's, an increase of ordering temperature above 200 K still remains a technological challenge.\cite{Dietl2010a} Rather surprisingly, the appearance of room temperature ferromagnetism in nonmagnetic electronic materials has been recently observed in many materials such as oxide and nitride dielectrics and semiconductors.\cite{Matsumoto2001a, Venkatesan2004a, Takeuchi2004a, Hong2006a, Yoon2006a} It has been proposed that defect, surface and localized states are the source of magnetic moments leading to high temperature ferromagnetism in nonmagnetic materials. \\ \indent The presence of acceptor-like defects has been proposed to be the likely sources of local magnetic moment in wide-gap semiconductors, such as GaN and ZnO.\cite{Dev2008a, Gohda2008a, Mitra2009a, Dev2010a} The acceptor-like defects can have important duality - they are not only fairly localized but also have extended tails that could allow for long-ranged magnetic coupling. Experimental investigations of the role of intrinsic defects in the magnetic ordering observed in wide band-gap semiconductors are scarce,\cite{Roul2011a, Roever2011a} as most of the research has been based on electronic structure calculations. Defect related magnetic phenomena is an important issue involved in high effective moment and room temperature ferromagnetism observed recently in GaN:Gd samples.\cite{Roever2011a} Further, in order to be certain of studying only the magnetic effects produced by intrinsic defects, they need to be introduced in a controlled way, e.g., through particle irradiation.\cite{Tuomisto2007a} In this work, we investigate the magnetic properties of 3$\;$$\mu$m-thick GaN layers grown by metal-organic chemical-vapor deposition (MOCVD) on sapphire substrates with point defects controllably created via high energy He irradiation, as measured by Tuomisto et al. in Ref. \onlinecite{Tuomisto2007b}. Three irradiated samples studied in this work cover wide range of concentrations of the introduced Ga vacancies in the range 10$^{17}$$-$10$^{19}$$\;$cm$^{-3}$, while the vacancy concentrations are below the detection limit (10$^{16}$$\;$cm$^{-3}$) of positron annihilation spectroscopy in the non-irradiated reference. \\ \indent The magnetic properties of the MOCVD GaN samples were studied with the use of Quantum Design MPMS XL-5 Squid Magnetometer. Measurements were carried out over broad temperature range of 5$-$300$\;$K with the use of constant magnetic field of induction up to $B$$\,$$\leq$$\,$5$\;$T.
\begin{figure}[b]
  \begin{center}
    \includegraphics[width = 0.40\textwidth, bb = 10 50 560 530]
    {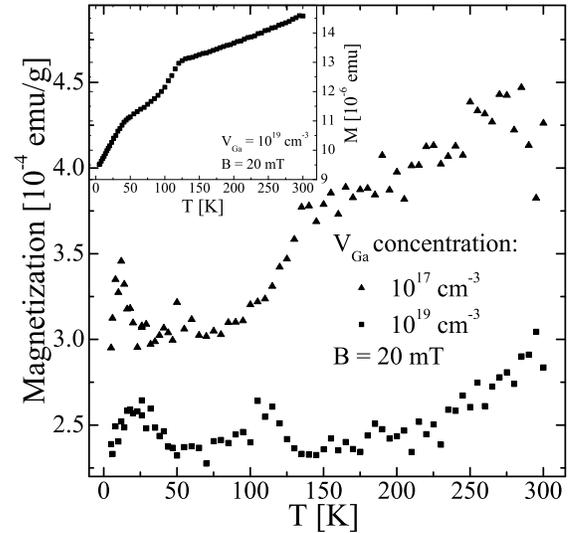}\\
  \end{center}
  \caption{\small The temperature dependence of magnetization measured at $B$$\,$$=$$\,$20$\;$mT for two He-irradiated GaN samples. The inset shows data obtained in a representative sample prior to substraction of the sapphire substrate effects.} \label{fig01}
\end{figure}
The measurements were performed in two steps. At first, both temperature and field dependence of magnetization $M$ was measured for all the studied layers. After the first sequence of measurements all GaN layers were etched with the use of inductively-coupled plasma reactor. Next, the magnetization of each individual sapphire substrate was measured separately in order to properly subtract its contribution from the overall signal. As can be seen in the inset to Fig.$\;$1 the magnetization of the GaN layer and the sapphire substrate is much larger and has a different shape than the shape of the $M$($T$) curve for GaN layer itself. In this way we could be sure that the results of the measurements were not corrupted by the possible magnetic impurities present in the substrate material. It should be also noted, that the control over magnetic purity of sapphire substrates is poor. \\ \indent The magnetic properties of the GaN layers were probed with the use of temperature dependent magnetization measurements. The measurements were performed with the use of low magnetic field $B$$\,$$=$$\,$20$\;$mT in the temperature range of 4.5$-$300$\;$K. It should be noted that the non-irradiated GaN reference samples showed weak, almost temperature independent diamagnetic response with magnetization $M$$\,$$\approx$$\,$$-$10$^{-6}$$\;$emu/g at $B$$\,$$=$$\,$20$\;$mT. The results obtained for the two irradiated samples are presented in Fig.$\:$\ref{fig01}. The magnetization for both studied samples showed positive, paramagnetic response with small changes for 4.5$\,$$\leq$$\,$$T$$\,$$\leq$$\,$300$\;$K. The lack of Curie behavior of the magnetization at low temperatures is a clear signature that the magnetic moments in the irradiated samples showed magnetic ordering with transition temperatures above 300$\;$K. The small changes of the magnetization with increasing temperature observed in the case of both irradiated samples cannot be attributed to any phase transition within the material. Significant, almost temperature independent magnetization $M$ at 4.5$\,$$\leq$$\,$$T$$\,$$\leq$$\,$300$\;$K indicate the presence of magnetic ordering with critical temperatures higher than 300$\;$K. The small increase of the $M$($T$) curves suggest that we do observe superparamagnetic behavior of magnetic defects with blocking temperature higher than room temperature. However, more detailed measurements needs to be done in order to determine the nature of the magnetic interactions between magnetic moments in the irradiated GaN samples. \\ \indent The same experimental setup was used to study isothermal magnetization hysteresis curves in  the irradiated GaN samples. All the hysteresis measurements were performed at $T$$\,$$=$$\,$5$\;$K. Before the measurement, the samples were cooled down from room temperature in the absence of external magnetic fields. The magnetic hysteresis curve was measured after magnetizing the sample up to the maximum field $B$$\,$$=$$\,$5$\;$T. The obtained magnetic hysteresis curves for GaN samples with both low and high gallium vacancy content are presented in Fig.$\:$\ref{fig02}.
\begin{figure}[t]
  \begin{center}
    \includegraphics[width = 0.40\textwidth, bb = 10 50 560 620]
    {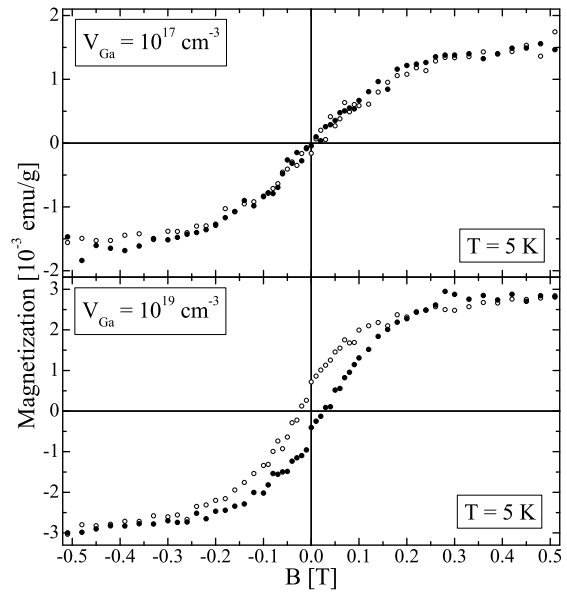}\\
  \end{center}
  \caption{\small Magnetic hysteresis loop measured at $T$$\,$$=$$\,$5$\;$K for two irradiated GaN samples.} \label{fig02}
\end{figure}
Irreversible magnetization was not observed in the case of GaN samples with lower concentration of vacancies, while the results clearly show magnetic hysteresis in the case of irradiated GaN sample with [$V_{Ga}$]$\,$$=$$\,$10$^{19}$$\;$cm$^{-3}$. The presence of magnetic hysteresis with measurable coercive field $H_{C}$$\,$$\approx$$\,$270$\;$Oe is a clear signature of the presence of magnetic irreversibility in this sample. The non-ellipsoidal shape of the magnetic hysteresis curve suggests that the level of magnetic disorder is rather small. It seems to be highly probable that in the case of high-vacancy content samples magnetic defects are aggregated into regions with sizes big enough enabling an appearance of multi-domain structure and thus explaining the appearance of the magnetic hysteresis curve. We interpret the change in magnetic behavior to originate from an increase in the concentration of magnetic moments in the studied samples. \\ \indent The magnetic behavior of the irradiated GaN samples in the presence of high magnetic fields was also investigated. We performed the magnetization measurements in the presence of static magnetic fields up to $B$$\,$$=$$\,$5$\;$T. Before proper measurement, each sample was cooled down in the absence of magnetic field. The magnetization was measured as a function of the applied magnetic field at constant temperature $T$$\,$$=$$\,$5$\;$K. The results for selected samples with different vacancy concentrations are shown in Fig.$\:$\ref{fig03}.
\begin{figure}[t]
  \begin{center}
    \includegraphics[width = 0.40\textwidth, bb = 5 40 570 570]
    {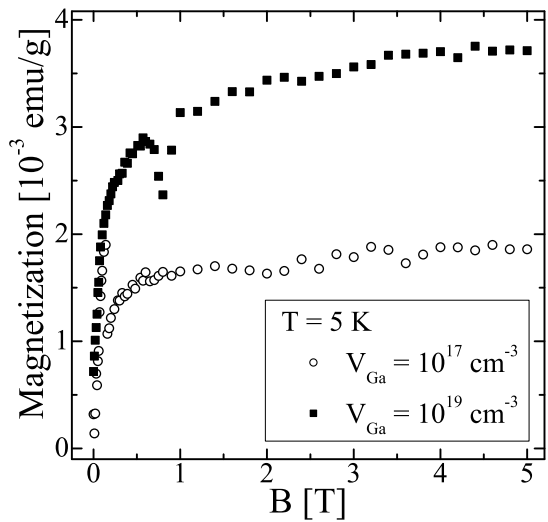}\\
  \end{center}
  \caption{\small Isothermal magnetization curves measured for the two selected He-irradiated GaN layers at $T$$\,$$=$$\,$5$\;$K.} \label{fig03}
\end{figure}
The magnetization rises more rapidly with the increase in the applied magnetic field for the sample showing a clear hysteresis curve. The observed behavior can be attributed to stronger magnetic interactions in the sample with the high concentration of defects. The magnetization $M(B)$ curve saturates for smaller magnetic fields in the case of the sample with less vacancies than in the ferromagnetic sample, indicating that the level of magnetic frustration is lower when the vacancy concentration is lower. This is in line with our interpretation of having more magnetic moments in the sample with higher vacancy concentration: the probability of finding two magnetic moments close to each other, leading to frustration, is higher when the concentration of magnetic moments is higher. However, the magnetization of all irradiated samples reached saturation at moderate magnetic fields $B$$\,$$<$$\,$4$\;$T indicating relatively homogeneous distribution of magnetic moments, because in the case of strong clustering the observed $M(B)$ curve should show tendency towards linear magnetic field dependence.\cite{Nagaev1995a} \\ \indent Both magnetization curves showed saturation with values of saturation magnetization $M_{S}$ equal to 1.9$\times$10$^{-3}$$\;$emu/g and 3.7$\times$10$^{-3}$$\;$emu/g for the samples with Ga vacancy concentrations of about 10$^{17}$$\;$cm$^{-3}$ and 10$^{19}$$\;$cm$^{-3}$, respectively. The observed magnetization curves show that the value of saturation magnetization is higher for the sample with higher vacancy concentration, while evidently the correspondence between the concentrations of the irradiation-induced defects and the magnetization is non-linear. The changes in the $M_{S}$ values observed in our experiments are much smaller than the differences between the vacancy concentrations, indicating that only a small fraction of defects created during the irradiation processes possessed nonzero magnetic moment. The amount of the magnetic moments present in each studied GaN sample can be determined from the value of the saturation magnetization, $M_{S}$. We used the following formula describing the saturation magnetization for the calculation of the amount of the magnetic moment concentration: $M_{S}$$\,$$=$$\,$$N_{A}$$\cdot$$J$$\cdot$$g$$\cdot$$\mu_{B}$$\cdot$$x$, where $N_{A}$ is the Avogadro constant, $J$$\,$$=$$\,$$L$$+$$S$ is the total magnetic moment consisting of orbital $L$ and spin moment $S$ of a single magnetic defect, $g$ is the spin splitting factor, $\mu_{B}$ is the Bohr magnetron, and $x$ is the concentration of magnetic defects. The estimation of magnetic moment concentration was done with the assumption that each magnetic defect only has a spin component of the magnetic moment $J$$\,$$=$$\,$$S$$\,$$=$$\,$3/2.\cite{Dev2010b} The estimated values of $x$ were equal to $x$$\,$$\approx$$\,$4.3$\times$10$^{17}$$\;$cm$^{-3}$ and $x$$\,$$\approx$$\,$8.3$\times$10$^{17}$$\;$cm$^{-3}$ for the samples with Ga vacancy concentrations of 10$^{17}$$\;$cm$^{-3}$ and 10$^{19}$$\;$cm$^{-3}$, respectively. It is clear that the rate at which magnetic moments are introduced by irradiation is different and non-linear compared to the Ga vacancy production. The as-grown GaN layers usually contain low concentration of Ga vacancy-related defects.\cite{Tuomisto2007a} Unintentional oxygen and hydrogen doping with concentrations around 10$^{17}$$\;$cm$^{-3}$ is typical of MOCVD-GaN.\cite{Hautakangas2006a} In-grown Ga-vacancies are usually complexed with oxygen impurities.\cite{Dev2008a} The small concentration of the observed magnetically-active defects with respect to the total Ga- vacancy concentration suggests that the presence of the oxygen/nitrogen-related vacancy complexes is the source of the observed magnetic moments. It has been shown theoretically that V$_{\textrm{Ga}}$-O$_{\textrm{N}}$ complexes do possess magnetic moment.\cite{Dev2008a} In principle, also interstitial-type defects are created in irradiation, but they are not stable at room temperature,\cite{Chow2004a} hence concentrating on the vacancy-type defects is justified. \\ \indent In conclusion, the presence of paramagnetic magnetic moments was observed in GaN with various concentrations of irradiation-induced Ga vacancies. The appearance of magnetic moments with concentration of about $x$$\,$$\approx$$\,$4.3$-$8.3$\times$10$^{17}$$\;$cm$^{-3}$ depending on the concentration of irradiation-induced defects. The magnetic defects show superparamagnetic behavior with blocking temperatures higher than 300$\;$K. An increase of the vacancy concentration resulted with appearance of well defined magnetic hysteresis at $T$$\,$$=$$\,$5$\;$K which is attributed to an increasing probability of vacancy aggregation into bigger entities. Our experimental results show that magnetic irreversibility can indeed arise from (irradiation-induced) intrinsic defects in GaN.

\begin{acknowledgments}

The samples were irradiated in the Lawrence Berkeley National Laboratory supported by US Department of Energy. The authors would like to thank C. Rauch for helpful discussions. Part of this research project has been supported by the European Commission under the 7th Framework Program through the Marie Curie Initial Training Network RAINBOW, Contract No. PITN-GA-2008-213238.

\end{acknowledgments}

\end{document}